\documentclass[journal]{IEEEtran}

\ifCLASSINFOpdf
\else
   \usepackage[dvips]{graphicx}
\fi
\usepackage{url}

\usepackage{graphicx}
\usepackage{amsmath,amssymb,amsfonts}
\usepackage{algorithmic}
\usepackage{algorithm}
\usepackage{cite}
\usepackage[colorlinks=true,linkcolor=blue,citecolor=blue,urlcolor=black]{hyperref}

\begin{document}

\title{Wasserstein Distributionally Robust Adaptive Beamforming}

\author{Kiarash~Hassas~Irani, Sergiy~A.~Vorobyov,~\IEEEmembership{Fellow,~IEEE}, and~Yongwei~Huang,~\IEEEmembership{Senior Member,~IEEE} 

\thanks{Kiarash~Hassas~Irani and Sergiy~A.~Vorobyov are with the Department of Information and Communications Engineering,
Aalto University, 02150 Espoo, Finland (e-mail: \href{mailto:kiarash.hassasirani@aalto.fi}{kiarash.hassasirani@aalto.fi}; \href{mailto:sergiy.vorobyov@aalto.fi}{sergiy.vorobyov@aalto.fi}).

Yongwei~Huang is with the School of Information Engineering, Guangdong University of Technology, University Town, Guangzhou, Guangdong 510006, China (e-mail: \href{mailto:ywhuang@gdut.edu.cn}{ywhuang@gdut.edu.cn}).}

\thanks{\textit{(Corresponding author: Sergiy~A.~Vorobyov.)}}
}

\maketitle

\begin{abstract}

Distributionally robust optimization (DRO)-based robust adaptive beamforming (RAB) enables enhanced robustness against model uncertainties, such as steering vector mismatches and interference-plus-noise covariance matrix estimation errors. Existing DRO-based RAB methods primarily rely on uncertainty sets characterized by the first- and second-order moments. In this work, we propose a novel Wasserstein DRO-based beamformer, using the worst-case signal-to-interference-plus-noise ratio maximization formulation. The proposed method leverages the Wasserstein metric to define uncertainty sets, offering a data-driven characterization of uncertainty. We show that the choice of the Wasserstein cost function plays a crucial role in shaping the resulting formulation, with norm-based and Mahalanobis-like quadratic costs recovering classical norm-constrained and ellipsoidal robust beamforming models, respectively. This insight highlights the Wasserstein DRO framework as a unifying approach, bridging deterministic and distributionally robust beamforming methodologies.

\end{abstract}


\IEEEpeerreviewmaketitle

\section{Introduction}

\IEEEPARstart{T}{he} primary challenge in robust adaptive beamforming (RAB) lies in uncertainties associated with the estimation of the steering vector and the interference-plus-noise covariance (INC) matrix, which can significantly degrade performance \cite{Li2005}. To address these challenges, several robust beamforming approaches have been developed. Among these, worst-case RAB methods, often framed within the minimum variance distortionless response (MVDR) formulation, have gained significant attention. These techniques aim to optimize the beamformer weights to preserve a distortionless response to the desired signal while minimizing the interference-plus-noise power at the array output, even under the most severe uncertainty conditions \cite{Vorobyov2003, Lorenz2005}. Another notable approach is the chance-constrained formulation, which aims to ensure that the probability of meeting the distortionless response constraint exceeds a specified threshold, thereby providing a probabilistic guarantee of robust performance \cite{Vorobyov2008}. Both of these methodologies have significantly advanced RAB design, with extensive research devoted to their theoretical foundations and optimization strategies \cite{Gershman2010, Vorobyov2013, Kim2008, ShahbazPanahi2018, Vorobyov2014, Elbir2023}.

Distributionally robust optimization (DRO) (see, e.g., \cite{Rahimian2022, Delage2024}) has emerged as a powerful framework in RAB, enabling robust performance without making strong deterministic assumptions about the underlying probability distributions of the steering vector and INC matrix. Unlike conventional RAB techniques, which typically define uncertainty sets in a fixed, deterministic manner, DRO-based methods construct these sets based on statistical moments or divergence measures, allowing for a more flexible, data-driven approach to uncertainty modeling. Within this framework, RAB optimization is performed against the least favorable distribution within the defined uncertainty set, ensuring reliable performance across all plausible distributions.

Chance-constrained DRO-based RAB formulations have been investigated in \cite{Zhang2016} and \cite{Li2018}. Worst-case DRO-based RAB formulations have been explored in \cite{Zhang2015} and \cite{Irani2025}. The study in \cite{Zhang2015} addresses the uncertainty in the first-order moment of the steering vector by ensuring that its expected value lies within an ellipsoidal region for all distributions belonging to the uncertainty set. A more comprehensive approach is taken in \cite{Irani2025}, where both the INC matrix and the steering vector are treated as uncertain parameters. The problem is formulated as a worst-case DRO with uncertainty sets defined based on first- and second-order moment constraints. 
 
Although uncertainty sets defined based on the statistical moments are computationally convenient, they may provide an incomplete representation of uncertainty, as they do not account for higher-order structural properties or the true shape of the distribution. This limitation can lead to sensitivity to moment estimation errors and reduced robustness in practical scenarios where data distributions exhibit complex variations.

In this work, we propose a novel DRO-based RAB formulation that leverages the Wasserstein metric to construct uncertainty sets for RAB under distributional uncertainty. Uncertainty sets defined via the Wasserstein distance offer a comprehensive and flexible approach by measuring the minimal transportation cost required to transform one probability distribution into another \cite{MohajerinEsfahani2018}. Unlike uncertainty sets defined based on the statistical moments, the Wasserstein metric inherently incorporates geometric information about the distribution, enabling a richer characterization of uncertainty.
By incorporating the geometric structure of probability distributions, Wasserstein-based formulations mitigate sensitivity to model misspecifications and estimation errors, resulting in beamforming designs that are inherently more robust. To exploit these advantages, the proposed method constructs uncertainty sets centered around empirical distributions, ensuring adaptability to data-driven scenarios. 

\section{Problem Formulation}

Consider an array of $N$ sensor elements. A point signal source $s(t)$, at the time instance $t$, impinges on the array with a steering vector $\boldsymbol{a}$. The received signal is modeled as
\begin{equation} \label{eq1}
    \boldsymbol{x}(t) = s(t)\boldsymbol{a} + \boldsymbol{i}(t) + \boldsymbol{n}(t),
\end{equation}
where $\boldsymbol{i}(t)$ and $\boldsymbol{n}(t)$ denote the complex-valued interference and noise vectors, respectively.

Applying the complex-valued beamforming weight vector $\boldsymbol{w}$, the beamformer output signal is given by
\begin{equation} \label{eq2}
    y(t) = \boldsymbol{w}^H \boldsymbol{x}(t),
\end{equation}
where $(\cdot)^H$ denotes the Hermitian transpose. Substituting \eqref{eq1} into \eqref{eq2} yields
\begin{equation} \label{eq3}
    y(t) = s(t)\boldsymbol{w}^H \boldsymbol{a} + \boldsymbol{w}^H (\boldsymbol{i}(t) + \boldsymbol{n}(t)).
\end{equation}

Based on \eqref{eq3}, the output signal-to-interference-plus-noise ratio (SINR) can be expressed as
\begin{equation}
    \text{SINR} = \frac{\sigma_{\rm s}^2 |\boldsymbol{w}^H \boldsymbol{a}|^2}{\boldsymbol{w}^H \boldsymbol{R}_{\rm{i+n}} \boldsymbol{w}},
\end{equation}
where $\sigma_{\rm s}^2$ represents the power of the desired signal, and $\boldsymbol{R}_{\rm{i+n}}$ is the INC matrix defined by
\begin{equation}
    \boldsymbol{R}_{\rm{i+n}} \triangleq \mathbb{E}[(\boldsymbol{i}(t) + \boldsymbol{n}(t))(\boldsymbol{i}(t) + \boldsymbol{n}(t))^H].
\end{equation}
In practice, the true INC matrix $\boldsymbol{R}_{\rm{i+n}}$ is usually unknown and is substituted by the sample data covariance matrix $\hat{\boldsymbol{R}} = \frac{1}{T}\sum_{t=1}^{T} \boldsymbol{x}(t)\boldsymbol{x}^H(t)$, where $T$ is the number of snapshots.

The objective in the classical MVDR beamforming design is to compute the weight vector $\boldsymbol{w}$ that maximizes the SINR while preserving a distortionless response to the desired signal. This can be formulated as the following optimization problem:
\begin{equation}
    \min_{\boldsymbol{w}} \; \boldsymbol{w}^H \hat{\boldsymbol{R}} \boldsymbol{w} \quad \text{s.t.} \; \ \boldsymbol{w}^H \boldsymbol{a} = 1.
\end{equation}
To account for possible mismatches between the presumed and actual steering vectors, the MVDR formulation relaxes the equality constraint to an inequality \cite{Vorobyov2003, Lorenz2005}:
\begin{equation}
    \min_{\boldsymbol{w}} \; \boldsymbol{w}^H \hat{\boldsymbol{R}} \boldsymbol{w} \quad \text{s.t.} \; \ \Re \{\boldsymbol{w}^H \boldsymbol{a}\} \geq 1.
\end{equation}

For ease of computation, we convert the above problem into its real-valued form by defining $\boldsymbol{w}_{\rm r} = [\Re \{\boldsymbol{w}\}^T \ \Im \{\boldsymbol{w}\}^T]^T$ and $\boldsymbol{a}_{\rm r} = [\Re \{\boldsymbol{a}\}^T \ \Im \{\boldsymbol{a}\}^T]^T$, where $\Re \{\cdot\}$ and $\Im \{\cdot\}$ denote the real and imaginary parts of a complex-valued vector, respectively.
Similarly, the real-valued representation of the INC matrix is given by
\begin{equation}
    \hat{\boldsymbol{R}}_{\rm r} = \begin{bmatrix} \Re \{\hat{\boldsymbol{R}}\} & -\Im \{\hat{\boldsymbol{R}}\} \\ \Im \{\hat{\boldsymbol{R}}\} & \Re \{\hat{\boldsymbol{R}}\} \end{bmatrix}.
\end{equation}
Thus, the MVDR problem becomes
\begin{equation}\label{RMVBProblem}
    \min_{\boldsymbol{w}_{\rm r}} \; \boldsymbol{w}_{\rm r}^T \hat{\boldsymbol{R}}_{\rm r} \boldsymbol{w}_{\rm r} \quad \text{s.t.} \; \ \boldsymbol{w}_{\rm r}^T \boldsymbol{a}_{\rm r} \geq 1.
\end{equation}

To account for uncertainty in the steering vector, we treat $\boldsymbol{a}_{\rm r} \in \mathbb{R}^{2N}$ as a random variable and reformulate \eqref{RMVBProblem} as
\begin{equation} \label{DRORAB1}
    \min_{\boldsymbol{w}_{\rm r}} \; \boldsymbol{w}_{\rm r}^T \hat{\boldsymbol{R}}_{\rm r} \boldsymbol{w}_{\rm r} \quad \text{s.t.} \; \ \inf_{P \in \mathcal{P}} \; \mathbb{E}_{\boldsymbol{a}_{\rm r} \sim P}[\boldsymbol{w}_{\rm r}^T \boldsymbol{a}_{\rm r}] \geq 1,
\end{equation}
where $\mathcal{P}$ denotes an uncertainty set for the probability distribution of $\boldsymbol{a}_{\rm r}$. Note that problem~\eqref{DRORAB1} can be equivalently expressed as
\begin{equation} \label{DRORAB2}
    \min_{\boldsymbol{w}_{\rm r}} \; \boldsymbol{w}_{\rm r}^T \hat{\boldsymbol{R}}_{\rm r} \boldsymbol{w}_{\rm r} \quad \text{s.t.} \; \ \sup_{P \in \mathcal{P}} \; \mathbb{E}_{\boldsymbol{a}_{\rm r} \sim P}[-\boldsymbol{w}_{\rm r}^T \boldsymbol{a}_{\rm r}] \leq -1,
\end{equation}

In our analysis, we characterize the uncertainty set $\mathcal{P}$ as a Wasserstein ball of radius $\epsilon$, centered at a nominal distribution $\widehat{P}$. Specifically,
\begin{equation}
    \mathcal{B}_\epsilon (\widehat{P}) = \left\{ P \in \mathcal{M}(\mathcal{X}) \; : \; W_\alpha(P, \widehat{P}) \leq \epsilon \right\}, 
\end{equation}
where $\mathcal{M}(\mathcal{X})$ is the set of all probability distributions with finite moments, supported on $\mathcal{X} = \mathbb{R}^{2N}$, and the \mbox{$\alpha$-Wasserstein} distance between two probability measures $P$ and $Q$, defined on a metric space $(\mathcal{X}, d)$, is given by
\begin{equation}
    W_\alpha(P, Q) = \left( \inf_{\gamma \in \Gamma(P,Q)} \int_{\mathcal{X} \times \mathcal{X}} d^\alpha(\boldsymbol{x}, \boldsymbol{y}) \, d\gamma(\boldsymbol{x}, \boldsymbol{y}) \right)^{1/\alpha},
\end{equation}
where $\Gamma(P,Q)$ denotes the set of all joint distributions with marginals $P$ and $Q$, and $d(\boldsymbol{x}, \boldsymbol{y})$ represents a metric between points $\boldsymbol{x}$ and $\boldsymbol{y}$ in $\mathcal{X}$. 

A data-driven choice for the nominal distribution is the empirical measure constructed from observed samples, given by
\begin{equation}
    \widehat{P}_M = \frac{1}{M} \sum_{i=1}^M \delta_{\hat{\boldsymbol{a}}_{\rm r}^i},
\end{equation}
where $M$ is the number of available samples, and $\delta_{\hat{\boldsymbol{a}}_{\rm r}^i}$ represents the Dirac point measure at the $i$\textsuperscript{th} observed sample $\hat{\boldsymbol{a}}_{\rm r}^i$.

\section{Main Results}

Using the strong duality property in Wasserstein DRO (see, e.g., \cite{Gao2023}), the dual formulation corresponding to the constraint in \eqref{DRORAB2} can be expressed as
\begin{equation} \label{dualDRORAB2}
    \min_{\lambda \geq 0}\left \{ \lambda \epsilon ^ \alpha + \mathbb{E}_{\hat{\boldsymbol{a}}_{\rm r} \sim \widehat{P}_M}\left[\sup_{\boldsymbol{a}_{\rm r} \in \mathbb{R}^{2N}} \left \{- \boldsymbol{w}_{\rm r}^T \boldsymbol{a}_{\rm r} - \lambda d^\alpha(\hat{\boldsymbol{a}}_{\rm r}, \boldsymbol{a}_{\rm r}) \right\}\right] \right \}.
\end{equation}

Let $\alpha = 1$ and the metric $d$ be the norm distance function. Then, the inner supremum in \eqref{dualDRORAB2} can be rewritten as
\begin{equation}\label{innerSup}
    \sup_{\boldsymbol{a}_{\rm r} \in \mathbb{R}^{2N}} \left \{ - \boldsymbol{w}_{\rm r}^T \boldsymbol{a}_{\rm r} - \lambda \left\|\hat{\boldsymbol{a}}_{\rm r} - \boldsymbol{a}_{\rm r}\right\| \right\}.
\end{equation}
Changing variable $\tilde{\boldsymbol{a}}_{\rm r} = \hat{\boldsymbol{a}}_{\rm r} - \boldsymbol{a}_{\rm r}$, \eqref{innerSup} can be reformulated in terms of $\tilde{\boldsymbol{a}}_{\rm r}$, and the supremum is equal to
\begin{equation}\label{innerSup2}
    - \boldsymbol{w}_{\rm r}^T \hat{\boldsymbol{a}}_{\rm r} + \sup_{\tilde{\boldsymbol{a}}_{\rm r} \in \mathbb{R}^{2N}} \left \{\boldsymbol{w}_{\rm r}^T \tilde{\boldsymbol{a}}_{\rm r} - \lambda \left\|\tilde{\boldsymbol{a}}_{\rm r}\right\| \right\}.
\end{equation}
The supremum in \eqref{innerSup2} corresponds to the convex conjugate of the function $f(\boldsymbol{x}) = \lambda \left\|\boldsymbol{x}\right\|$ for $\lambda > 0$ (note that for $\lambda = 0$, the supremum is $+\infty$), evaluated at $\boldsymbol{w}_{\rm r}$, i.e., $f^*(\boldsymbol{w}_{\rm r})$. Exploiting the scaling property of conjugate functions and the fact that the conjugate of a norm function results in an indicator function over the corresponding dual norm unit ball \cite{Boyd2004}, we have
\begin{equation}
    f^*(\boldsymbol{w}_{\rm r}) = \begin{cases}
        0, &\left\|\boldsymbol{w}_{\rm r}\right\|_* \leq \lambda\\
        +\infty, &\left\|\boldsymbol{w}_{\rm r}\right\|_* > \lambda,
    \end{cases}
\end{equation}
where $\left\|\cdot\right\|_*$ denotes the dual norm. 

Therefore, \eqref{dualDRORAB2} can be simplified as
\begin{equation} \label{dualDRORAB3}
    \min_{\lambda \geq \left\|\boldsymbol{w}_{\rm r}\right\|_*} \left \{ \lambda \epsilon + \mathbb{E}_{\hat{\boldsymbol{a}}_{\rm r} \sim \widehat{P}_M}\left[- \boldsymbol{w}_{\rm r}^T \hat{\boldsymbol{a}}_{\rm r}\right] \right \} = \epsilon \left\|\boldsymbol{w}_{\rm r}\right\|_* - \boldsymbol{w}_{\rm r}^T \bar{\boldsymbol{a}}_{\rm r},
\end{equation}
where $\bar{\boldsymbol{a}}_{\rm r}$ is the empirical mean of the real-valued representation of the steering vector samples.

Note that the dual problem \eqref{dualDRORAB3} remains unchanged if the empirical distribution $\widehat{P}_M$ is replaced with any other distribution with the same mean. This is because the expectation term in the constraint of \eqref{DRORAB2} only depends on the first-order moment of the random variable. To formalize this reasoning, we invoke the Kantorovich-Rubinstein theorem (see, e.g., \cite{Gao2023}), which states that for any Lipschitz function $\Psi(\cdot)$ with Lipschitz constant \textit{L}, the following inequality holds:
\begin{equation} \label{Kantorovich–Rubinstein}
    \left| \mathbb{E}_{P}\left[ \Psi(\cdot) \right] - \mathbb{E}_{\widehat{P}_M}\left[ \Psi(\cdot) \right] \right| \leq \textit{L} \, W_1(P, \widehat{P}_M) \leq \textit{L} \, \epsilon. 
\end{equation}
In our problem, the function of interest is $\Psi(\boldsymbol{a}_{\rm r}) = -\boldsymbol{w}_{\rm r}^T \boldsymbol{a}_{\rm r}$, for which the Lipschitz constant is given by $\textit{L} = \left\|\boldsymbol{w}_{\rm r}\right\|_*$. Consequently, the expectation term in the constraint of \eqref{DRORAB2} can be upper bounded as follows:
\begin{equation}
    \mathbb{E}_{\boldsymbol{a}_{\rm r} \sim P}\left[ -\boldsymbol{w}_{\rm r}^T \boldsymbol{a}_{\rm r} \right] \leq \epsilon \left\|\boldsymbol{w}_{\rm r}\right\|_* - \boldsymbol{w}_{\rm r}^T\mathbb{E}_{\boldsymbol{a}_{\rm r} \sim \widehat{P}_M}\left[ \boldsymbol{a}_{\rm r} \right].
\end{equation}
This upper bound depends solely on the first-order moment of the nominal distribution. Moreover, it implies that the dual variable $\lambda = \left\|\boldsymbol{w}_{\rm r}\right\|_*$ (it coincides here with \textit{L}) serves as a \emph{certificate of robustness}, that is, it guarantees that no other distribution within the defined Wasserstein ball can yield an expected value that exceeds the derived upper bound. Hence, the value of $\lambda$ quantifies the maximum deviation induced by distributional uncertainty, certifying that the solution remains robust against all admissible perturbations.

Consequently, the dual formulation of the Wasserstein DRO-based RAB problem~\eqref{DRORAB1} is given by
\begin{equation} \label{DRORAB3}
    \min_{\boldsymbol{w}_{\rm r}} \; \boldsymbol{w}_{\rm r}^T \hat{\boldsymbol{R}}_{\rm r} \boldsymbol{w}_{\rm r} \quad \text{s.t.} \; \ \epsilon \left\|\boldsymbol{w}_{\rm r}\right\|_* \leq \boldsymbol{w}_{\rm r}^T \bar{\boldsymbol{a}}_{\rm r} - 1.
\end{equation}

Specifically, we consider the case of using the Euclidean norm. Owing to the self-duality property of the Euclidean norm, problem~\eqref{DRORAB3} simplifies to a second-order cone program (SOCP), which is given by
\begin{equation} \label{DRORAB4}
    \min_{\boldsymbol{w}_{\rm r}} \; \boldsymbol{w}_{\rm r}^T \hat{\boldsymbol{R}}_{\rm r} \boldsymbol{w}_{\rm r} \quad \text{s.t.} \; \ \epsilon \left\|\boldsymbol{w}_{\rm r}\right\|_2 \leq \boldsymbol{w}_{\rm r}^T \bar{\boldsymbol{a}}_{\rm r} - 1.
\end{equation}

Based on the Lagrangian formulation of problem~\eqref{DRORAB4}, if the Lagrange multiplier associated with the inequality constraint were zero, the first-order optimality condition would enforce $\boldsymbol{w}_{\rm r} = \boldsymbol{0}$, which contradicts the feasibility condition. Therefore, by the complementary slackness condition, the inequality constraint must be active at optimality. This leads to the following equivalent reformulation:
\begin{equation} \label{DRORAB5}
    \min_{\boldsymbol{w}_{\rm r}} \; \boldsymbol{w}_{\rm r}^T \hat{\boldsymbol{R}}_{\rm r} \boldsymbol{w}_{\rm r} \quad \text{s.t.} \; \ \epsilon \left\|\boldsymbol{w}_{\rm r}\right\|_2 = \boldsymbol{w}_{\rm r}^T \bar{\boldsymbol{a}}_{\rm r} - 1.
\end{equation}

The constraint in \eqref{DRORAB5} implies that $\boldsymbol{w}_{\rm r}^T \bar{\boldsymbol{a}}_{\rm r} > \epsilon \left\|\boldsymbol{w}_{\rm r}\right\|_2$. Furthermore, applying the Cauchy–Schwarz inequality, we obtain $\boldsymbol{w}_{\rm r}^T \bar{\boldsymbol{a}}_{\rm r} \leq \left\|\boldsymbol{w}_{\rm r}\right\|_2 \left\|\bar{\boldsymbol{a}}_{\rm r}\right\|_2$. Combining these two inequalities, the feasibility is ensured if
\begin{equation}
    \epsilon < \left\|\bar{\boldsymbol{a}}_{\rm r}\right\|_2. 
\end{equation}

It is worth highlighting that the Wasserstein DRO-based RAB formulation with the Euclidean norm distance exhibits a strong analogy to the classical worst-case robust beamforming method proposed in \cite{Vorobyov2003}. A key distinction lies in the modeling of the steering vector: our approach treats the steering vector as a random variable, whereas \cite{Vorobyov2003} employs a deterministic additive mismatch model based on the presumed steering vector. Specifically, $\bar{\boldsymbol{a}}_{\rm r}$ in our formulation serves the same role as the real-valued representation of the presumed steering vector in \cite{Vorobyov2003}, while the parameter $\epsilon$ directly corresponds to the upper bound on the Euclidean norm of the real-valued representation of the steering vector mismatch. 

To rigorously establish this correspondence, we leverage the Kantorovich–Rubinstein theorem stated in \eqref{Kantorovich–Rubinstein}. Specifically, we consider the true distribution $P$ of the random vector $\boldsymbol{a}_{\rm r}$, whose first-order moment corresponds to the real-valued representation of the actual steering vector, i.e., $\mathbb{E}_{\boldsymbol{a}_{\rm r} \sim P}[\boldsymbol{a}_{\rm r}] = \boldsymbol{a}_{\rm r}^{\text{actual}}$. The real-valued representation of the presumed steering vector is then set as the empirical mean $\bar{\boldsymbol{a}}_{\rm r}$ of the samples. Defining the function $\Psi_{\boldsymbol{u}}(\boldsymbol{a}_{\rm r}) = \boldsymbol{u}^T (\boldsymbol{a}_{\rm r} - \bar{\boldsymbol{a}}_{\rm r})$ for any unit vector $\boldsymbol{u}$ (i.e., $\|\boldsymbol{u}\|_2 = 1$), we note that $\Psi_{\boldsymbol{u}}$ is a Lipschitz function with a Lipschitz constant of $1$ with respect to the Euclidean norm. Furthermore, its empirical expectation satisfies $\mathbb{E}_{\boldsymbol{a}_{\rm r} \sim \widehat{P}_M}\left[ \Psi_{\boldsymbol{u}}(\boldsymbol{a}_{\rm r}) \right] = 0$. Applying \eqref{Kantorovich–Rubinstein}, we obtain 
\begin{equation}
    \left| \mathbb{E}_{\boldsymbol{a}_{\rm r} \sim P}\left[ \Psi_{\boldsymbol{u}}(\boldsymbol{a}_{\rm r}) \right] \right| \leq \epsilon. 
\end{equation}
Taking the supremum over all unit vectors $\boldsymbol{u}$, it follows that
\begin{equation}
    \left\| \boldsymbol{a}_{\rm r}^{\text{actual}} - \bar{\boldsymbol{a}}_{\rm r} \right\|_2 \leq \epsilon. 
\end{equation}
Thus, the Wasserstein radius $\epsilon$ provides an upper bound on the Euclidean norm of the mismatch between the real-valued representations of the actual steering vector and its empirical estimate. This structural similarity suggests that the classical worst-case formulation in \cite{Vorobyov2003} implicitly accounts for a form of distributional uncertainty, despite being derived from a purely deterministic perspective.

Next, we analyze the problem using a different metric for the $1$-Wasserstein distance. To this end, we adopt a strictly convex quadratic function as a Mahalanobis-like metric, given by  
\begin{equation}
    d(\boldsymbol{x}, \boldsymbol{y}) = \frac{1}{2} (\boldsymbol{x} - \boldsymbol{y})^T \boldsymbol{\Lambda} (\boldsymbol{x} - \boldsymbol{y}),
\end{equation}  
where $\boldsymbol{\Lambda}$ is a positive definite matrix. In this case, the inner supremum in \eqref{dualDRORAB2} simplifies to
\begin{equation}\label{innerSup2_quadratic}
    - \boldsymbol{w}_{\rm r}^T \hat{\boldsymbol{a}}_{\rm r} + \sup_{\tilde{\boldsymbol{a}}_{\rm r} \in \mathbb{R}^{2N}} \left \{\boldsymbol{w}_{\rm r}^T \tilde{\boldsymbol{a}}_{\rm r} - \frac{\lambda}{2} \tilde{\boldsymbol{a}}_{\rm r}^T \boldsymbol{\Lambda} \tilde{\boldsymbol{a}}_{\rm r} \right\}.
\end{equation}
Here, the supremum corresponds to the convex conjugate of the quadratic function $g(\boldsymbol{x}) = \frac{\lambda}{2} \boldsymbol{x}^T \boldsymbol{\Lambda} \boldsymbol{x}$, evaluated at $\boldsymbol{w}_{\rm r}$, which is given by \cite{Boyd2004}:
\begin{equation}
    g^*(\boldsymbol{w}_{\rm r}) = \frac{1}{2\lambda} \boldsymbol{w}_{\rm r}^T \boldsymbol{\Lambda}^{-1} \boldsymbol{w}_{\rm r}.
\end{equation}
Accordingly, the dual problem takes the form
\begin{equation} \label{dual_quadratic}
    - \boldsymbol{w}_{\rm r}^T \bar{\boldsymbol{a}}_{\rm r} + \min_{\lambda \geq 0} \left \{ \lambda \epsilon + \frac{1}{2\lambda} \boldsymbol{w}_{\rm r}^T \boldsymbol{\Lambda}^{-1} \boldsymbol{w}_{\rm r} \right \}.
\end{equation}

The objective function in \eqref{dual_quadratic} is convex with respect to $\lambda$. Its minimizer $\lambda ^*$ is obtained by setting its derivative to zero and can be found as
\begin{equation} \label{lambdaStar}
\begin{split}
    \lambda^* =\sqrt{\frac{\boldsymbol{w}_{\rm r}^T \boldsymbol{\Lambda}^{-1} \boldsymbol{w}_{\rm r}}{2\epsilon}}. 
\end{split}   
\end{equation}
Substituting \eqref{lambdaStar} into \eqref{dual_quadratic}, the optimal dual value simplifies to
\begin{equation}
    - \boldsymbol{w}_{\rm r}^T \bar{\boldsymbol{a}}_{\rm r} + \sqrt{2\epsilon(\boldsymbol{w}_{\rm r}^T \boldsymbol{\Lambda}^{-1} \boldsymbol{w}_{\rm r})} = - \boldsymbol{w}_{\rm r}^T \bar{\boldsymbol{a}}_{\rm r} + \sqrt{2\epsilon} \left\| \boldsymbol{\Gamma}^{-1} \boldsymbol{w}_{\rm r} \right\|_2, 
\end{equation}
where $\boldsymbol{\Gamma}$ is the square root of the positive definite matrix $\boldsymbol{\Lambda}$, i.e., $\boldsymbol{\Lambda} = \boldsymbol{\Gamma}^2$. Consequently, the dual formulation of the Wasserstein DRO-based RAB problem in \eqref{DRORAB1} is given by
\begin{equation} \label{DRORAB_quadratic_2}
    \min_{\boldsymbol{w}_{\rm r}} \; \boldsymbol{w}_{\rm r}^T \hat{\boldsymbol{R}}_{\rm r} \boldsymbol{w}_{\rm r} \quad \text{s.t.} \; \ \sqrt{2\epsilon}\left\| \boldsymbol{\Gamma}^{-1} \boldsymbol{w}_{\rm r}\right\|_2 = \boldsymbol{w}_{\rm r}^T \bar{\boldsymbol{a}}_{\rm r} - 1,
\end{equation}
where the inequality constraint is replaced with an equality, as it is active at optimality.

The formulation in \eqref{DRORAB_quadratic_2} closely resembles the beamformer design proposed in \cite{Lorenz2005}, which considers an ellipsoidal uncertainty model for the steering vector. Specifically, the uncertainty set in \cite{Lorenz2005} takes the form
\begin{equation} \label{ellipsoid}
    \mathcal{E}(\bar{\boldsymbol{a}}_{\rm r}, 2\epsilon \boldsymbol{\Lambda}^{-1}) = \left\{ \boldsymbol{a}_{\rm r} \, \middle| \, (\boldsymbol{a}_{\rm r} - \bar{\boldsymbol{a}}_{\rm r})^T\boldsymbol{\Lambda}(\boldsymbol{a}_{\rm r} - \bar{\boldsymbol{a}}_{\rm r}) \leq 2\epsilon  \right\},
\end{equation} 
where the matrix $\boldsymbol{\Lambda}$ defines the shape of the ellipsoid (e.g., it can be set to the inverse of the sample data covariance matrix), $2\epsilon$ scales its size, and empirical mean $\bar{\boldsymbol{a}}_{\rm r}$ represents the center.

This observation underscores a fundamental connection between the metric employed in the Wasserstein DRO-based RAB formulation and the structure of the uncertainty set in deterministic robust models. In particular, when a norm-based metric is used, the resulting formulation coincides with that of the deterministic model defined via a norm constraint. Similarly, adopting a Mahalanobis-like quadratic metric yields a formulation that mirrors the deterministic model based on ellipsoidal uncertainty---where the uncertainty set is naturally characterized by a Mahalanobis distance. In this way, the deterministic robust approaches can be interpreted as special cases of the more general distributionally robust framework, with the choice of metric reflecting the geometry of the corresponding uncertainty set.

Assuming that the random steering vector $\boldsymbol{a}_{\rm r}$ follows a multivariate Gaussian distribution and that the matrix $\boldsymbol{\Lambda}$ is set to the inverse of the sample data covariance matrix, we can establish a connection between \eqref{DRORAB_quadratic_2} and the chance-constrained formulation of the problem in \eqref{RMVBProblem}. 
In particular, by choosing
\begin{equation}
    \epsilon = \frac{1}{2} \mathcal{\chi}_{2N, \beta}^2,
\end{equation}
where $\mathcal{\chi}_{2N, \beta}^2$ denotes the $\beta$-quantile function of the chi-squared distribution with $2N$ degrees of freedom, the ellipsoid in \eqref{ellipsoid} represents a confidence region that contains the true steering vector with probability at least $\beta$. Enforcing the distortionless response constraint to hold for all vectors within this ellipsoid guarantees that the following chance constraint is required to be satisfied:
\begin{equation}
    \mathbb{P}_{\boldsymbol{a}_{\rm r} \sim G}(\boldsymbol{w}_{\rm r}^T \boldsymbol{a}_{\rm r} \geq 1) \geq \mathbb{P}_{\boldsymbol{a}_{\rm r} \sim G} (\boldsymbol{a}_{\rm r} \in \mathcal{E}) = \beta,
\end{equation}
where $G$ denotes the Gaussian distribution.

The analysis thus far has focused on the distributional uncertainty in the steering vector. We now extend this framework to incorporate uncertainty in the INC matrix. It is also a critical factor that can significantly impact performance even when the steering vector is precisely known. Specifically, the DRO-based formulation corresponding to the objective in \eqref{RMVBProblem} is given by
\begin{equation} \label{DRORAB_INC}
    \min_{\boldsymbol{w}_{\rm r}} \; \sup_{Q \in \mathcal{B}_\rho (\widehat{Q})} \; \mathbb{E}_{\boldsymbol{R}_{\rm r} \sim Q}[\boldsymbol{w}_{\rm r}^T \boldsymbol{R}_{\rm r} \boldsymbol{w}_{\rm r}],
\end{equation}
where $\boldsymbol{R}_{\rm r} \succeq \boldsymbol{0}$ is the random INC matrix, and the uncertainty set is defined as a $1$-Wasserstein ball of radius $\rho$, centered at a nominal distribution $\widehat{Q}$. The latter is chosen such that the expected value of $\boldsymbol{R}_{\rm r}$ under $\widehat{Q}$ matches the sample data covariance matrix, i.e., $\mathbb{E}_{\boldsymbol{R}_{\rm r} \sim \widehat{Q}}[\boldsymbol{R}_{\rm r}] = \hat{\boldsymbol{R}}_{\rm r}$. To proceed, we adopt the Frobenius norm as the underlying metric, leading to the following dual formulation for the inner supremum:
\begin{equation} \label{dualDRORAB_INC}
    \min_{\lambda \geq 0}\left \{ \lambda \rho + \mathbb{E}_{\check{\boldsymbol{R}}_{\rm r} \sim \widehat{Q}}\left[\sup_{\boldsymbol{R}_{\rm r}} \left\{\boldsymbol{w}_{\rm r}^T \boldsymbol{R}_{\rm r} \boldsymbol{w}_{\rm r} - \lambda \| \boldsymbol{R}_{\rm r} - \check{\boldsymbol{R}}_{\rm r} \|_F \right\}\right] \right \}.
\end{equation}
We define $\tilde{\boldsymbol{R}}_{\rm r} = \boldsymbol{R}_{\rm r} - \check{\boldsymbol{R}}_{\rm r}$, and rewrite the supremum in \eqref{dualDRORAB_INC} as:
\begin{equation}
    \boldsymbol{w}_{\rm r}^T \check{\boldsymbol{R}}_{\rm r} \boldsymbol{w}_{\rm r} + \sup_{\tilde{\boldsymbol{R}}_{\rm r}} \left\{ \langle \tilde{\boldsymbol{R}}_{\rm r}, \boldsymbol{w}_{\rm r} \boldsymbol{w}_{\rm r}^T \rangle - \lambda \| \tilde{\boldsymbol{R}}_{\rm r} \|_F \right\}.
\end{equation}

Using the Cauchy-Schwarz inequality for the Frobenius inner product, the supremum can be evaluated as follows:
\begin{equation}
    \sup_{\tilde{\boldsymbol{R}}_{\rm r}} \left\{ \langle \tilde{\boldsymbol{R}}_{\rm r}, \boldsymbol{w}_{\rm r} \boldsymbol{w}_{\rm r}^T \rangle - \lambda \| \tilde{\boldsymbol{R}}_{\rm r} \|_F \right\} = \begin{cases}
        0,\! &\left\| \boldsymbol{w}_{\rm r} \boldsymbol{w}_{\rm r}^T \right\|_F \leq \lambda\\
        +\infty,\! &\left\| \boldsymbol{w}_{\rm r} \boldsymbol{w}_{\rm r}^T \right\|_F > \lambda.
    \end{cases}
\end{equation}
Note that $\left\| \boldsymbol{w}_{\rm r} \boldsymbol{w}_{\rm r}^T \right\|_F = \left\| \boldsymbol{w}_{\rm r} \right\|_2^2$. Consequently, the dual problem simplifies to
\begin{equation} \label{dualDRORAB_INC_2}
    \min_{\lambda \geq \|\boldsymbol{w}_{\rm r}\|_2^2} \left\{ \lambda \rho + \mathbb{E}_{\check{\boldsymbol{R}}_{\rm r} \sim \widehat{Q}}\left[\boldsymbol{w}_{\rm r}^T \check{\boldsymbol{R}}_{\rm r} \boldsymbol{w}_{\rm r}\right] \right\} = \boldsymbol{w}_{\rm r}^T \hat{\boldsymbol{R}}_{\rm r} \boldsymbol{w}_{\rm r} + \rho \|\boldsymbol{w}_{\rm r}\|_2^2.
\end{equation}

Therefore, problem \eqref{DRORAB_INC} is equivalent to
\begin{equation}
    \min_{\boldsymbol{w}_{\rm r}} \; \left\{ \boldsymbol{w}_{\rm r}^T \hat{\boldsymbol{R}}_{\rm r} \boldsymbol{w}_{\rm r} + \rho \|\boldsymbol{w}_{\rm r}\|_2^2 = \boldsymbol{w}_{\rm r}^T (\hat{\boldsymbol{R}}_{\rm r} + \rho \boldsymbol{I}) \boldsymbol{w}_{\rm r} \right\}.
\end{equation}

This result suggests that introducing the Wasserstein uncertainty in the INC matrix induces a regularization effect, effectively approximating the INC matrix as a diagonally loaded version of the sample data covariance matrix.

\section{Conclusion}

In this paper, we have introduced a novel Wasserstein DRO-based RAB formulation designed to ensure robustness against distributional uncertainties in beamforming. Unlike conventional DRO-based RAB approaches, which rely solely on the first- and second-order statistical moments, the proposed method constructs uncertainty sets using the \mbox{$1$-Wasserstein} metric, allowing for a comprehensive representation of uncertainty. We have demonstrated that the choice of metric in the Wasserstein DRO framework fundamentally shapes the resulting RAB formulation, revealing a deep connection between distributional and deterministic robustness. This insight highlights the Wasserstein DRO framework as a unifying one, where deterministic robust formulations emerge as special cases under appropriately chosen Wasserstein cost functions.

\end{document}